\documentstyle[epsfig]{aipproc}

\newcommand{\nc}{\newcommand}
\nc{\rnc}{\renewcommand}
\nc{\acs}{\arraycolsep}
\nc{\mc}{\multicolumn}
\nc{\bsk}{\baselineskip}
\nc{\vsp}{\vspace}
\nc{\hsp}{\hspace}
\nc{\stl}{\setlength}
\nc{\stc}{\setcounter}
\nc{\addl}{\addtolength}
\nc{\beq}{\begin{equation}}
\nc{\eeq}{\end{equation}}
\nc{\beqa}{\begin{eqnarray}}
\nc{\eeqa}{\end{eqnarray}}
\nc{\tfrac}[2]{\raisebox{.4ex}{\tiny $\frac{#1}{#2}$}} 
\nc{\romlist}{ \setcounter{num1}{0}%
  \begin{list}{(\roman{num1})}{\usecounter{num1}} }
\nc{\arblist}{ \setcounter{num1}{0}%
  \begin{list}{(\arabic{num1})}{\usecounter{num1}} }
\nc{\alphlist}{ \setcounter{num2}{0}%
  \begin{list}{(\alph{num2})}{\usecounter{num2}} }
\nc{\bullist}{\begin{list}{$\bullet$}{ }}
\nc{\nr}{\\ \hline}
\nc{\hrl}{{\center \stl{\unitlength}{\textwidth} 
 \begin{picture}(1,0)  \put(0,0){\line(1,0){1}}
 \end{picture} \vsp{.001\bsk} }}
\nc{\cents}{{\scriptsize$\mbox{\rm C}\!\!\!\mbox{\raisebox{.2ex}%
{$|$}}\,\,\,\,$}}
\nc{\figsp}[5]{\begin{figure}[#1] \vsp{#2} \caption[#4]{#3} 
\label{#5} \vsp{2\bsk} \end{figure}}
\nc{\fig}{\figsp{tbp}}
\nc{\figb}{\figsp{b}}
\nc{\figh}{\figsp{h}}
\nc{\llist}{\begin{list}{}{} \stl{\labelsep}{.4in}}
\nc{\lit}[2]{
 \item[\raggedright #1]{#2}}
\nc{\lbit}[2]{ 
 \item[\raggedright\bf #1]{#2}}
\nc{\lemit}[2]{ 
 \item[\raggedright\em #1]{#2}}
\nc{\lbemit}[2]{ 
 \item[\raggedright\bf\em #1]{#2}}
\nc{\clst}[1]{\stl{\coltwo}{\textwidth}
\addl{\coltwo}{-#1} \addl{\coltwo}{-5.56ex} \newline
\begin{tabular}{p{#1}p{\coltwo}} \citem{}{}}
 
\nc{\citem}[2]{{\raggedright \bf #1} & #2 \\ }
\nc{\cemitem}[2]{{\raggedright \em #1} & #2 \\ }
\nc{\cbemitem}[2]{{\raggedright \bf \em #1} & #2 \\ }
\nc{\cend}{\citem{}{} \end{tabular} 
\mbox{}
} 
\nc{\SSP}{{\rm \hsp{.4in}}}
\nc{\SSPP}{{\rm \hsp{.2in}}}
\nc{\ds}{\displaystyle}
\nc{\tx}{\textstyle}
\nc{\scst}{\scriptstyle}
\nc{\sscst}{\scriptscriptstyle}
\nc{\prt}{\partial}
\nc{\fr}{\frac}
\nc{\lf}{\left}
\nc{\rt}{\right}
\nc{\la}{\langle}
\nc{\ra}{\rangle}
\nc{\V}{\vec}
\nc{\str}{\stackrel}
\nc{\ovl}{\overline}
\nc{\ul}{\underline}
\nc{\ovb}{\overbrace}
\nc{\wh}{\widehat}
\nc{\B}{\bar}
\nc{\D}{\dot}
\nc{\C}{\cdot}
\nc{\dd}{\ddot}
\nc{\tl}{\tilde}
\nc{\ha}{\hat}
\nc{\nn}{\nonumber}
\nc{\app}{\approx}
\nc{\al}{\alpha}
\nc{\RA}{\rightarrow}
\nc{\LRA}{\leftrightarrow}
\nc{\SRA}{\SSP\rightarrow\SSP}
\nc{\SSRA}{\SSPP\rightarrow\SSPP}
\nc{\dg}{\dagger}
\nc{\vp}{\varphi}
\nc{\ve}{\varepsilon}
\nc{\Dl}{\Delta}
\nc{\dl}{\delta}
\nc{\gm}{\gamma}
\nc{\Gm}{\Gamma}
\nc{\ep}{\epsilon}
\nc{\sg}{\sigma}
\nc{\Sg}{\Sigma}
\nc{\ua}{\uparrow}
\nc{\da}{\downarrow}
\nc{\lam}{\lambda}
\nc{\eql}[1]{\parbox{#1\textwidth}}
\nc{\eqm}[1]{\makebox[#1\textwidth][l]}
\nc{\enu}[1]{\mbox{\hspace{.4in}(\theequation.#1)}}
\nc{\son}{\\ \\ \ds}
\nc{\stw}{\\ & \\ \ds}		
\nc{\sth}{\\ & & \\ \ds}
\nc{\sfo}{\\ & & & \\ \ds}
\nc{\sfi}{\\ & & & & \\ \ds}
\nc{\A}{& \ds}
\nc{\bbr}{\lf\{\rule[-1.5ex]{0in}{0.01in}\rt.}
\nc{\hf}{\fr{1}{2}}
\nc{\mhf}{\mbox{\footnotesize$\hf$}}
\nc{\dv}{\/!}
\nc{\dint}{\int\!\!\int}
\nc{\tint}{\int\!\!\dint}               
\nc{\qint}{\int\!\!\tint}
\nc{\Pd}[2]{\fr{\prt #1}{\prt #2}}
\nc{\Pdt}[1]{\Pd{#1}{t}}
\nc{\Pdx}[1]{\Pd{#1}{x}}
\nc{\Pdy}[1]{\Pd{#1}{y}}
\nc{\Pdz}[1]{\Pd{#1}{z}}           	
\nc{\Pdr}[1]{\Pd{#1}{r}}
\nc{\Pds}[1]{\Pd{#1}{s}}
\nc{\Dv}[2]{\fr{d#1}{d#2}}
\nc{\Dvt}[1]{\Dv{#1}{t}}
\nc{\Dvx}[1]{\Dv{#1}{x}}
\nc{\Dvy}[1]{\Dv{#1}{y}}
\nc{\Dvz}[1]{\Dv{#1}{z}}
\nc{\Dvr}[1]{\Dv{#1}{r}}
\nc{\Drs}[1]{\Dv{#1}{s}}
\nc{\inpp}[3]{\la #1| #2| #3\ra}
\nc{\inp}[2]{\inpp{#1}{#2}{#1}}
\nc{\rb}[1]{| #1\ra}
\nc{\lb}[1]{\la#1|}
\nc{\dtpp}[2]{\lb{#1}\rb{#2}}		
\nc{\dtp}[1]{\dtpp{#1}{#1}}
\nc{\otpp}[2]{\rb{#1}\lb{#2}}
\nc{\otp}[1]{\otpp{#1}{#1}}

\newcounter{num1} \newcounter{num2}  
\newlength{\coltwo}
\rnc{\L}{{\cal L}}                      
\nc{\lapp}{\mbox{\raisebox{-.6ex}{$\,\stackrel{\textstyle <}{\sim}\,$}}}
\nc{\gapp}{\mbox{\raisebox{-.6ex}{$\,\stackrel{\textstyle >}{\sim}\,$}}}
\nc{\als}{\fr{\al_s(Q^2)}{2\pi}}
\nc{\gpx}{g_1^p(x,Q^2)}
\nc{\gpz}{g_1^p(z,Q^2)}
\nc{\muq}{\lf(\fr{\mu^2}{Q^2}\rt)}
\nc{\xy}{(\fr{x}{y})}
\nc{\ASQ}{\al_s(Q^2)}
\nc{\Li}{{\rm Li}_2}
\nc{\dqx}{\Dl q_i(x,Q^2)} \nc{\dqy}{\Dl q_i(y,Q^2)} 
\nc{\dQ}{\Dl q_i(Q^2)}
\nc{\dgx}{\Dl g(x,Q^2)} \nc{\dgy}{\Dl g(y,Q^2)} 
\nc{\dG}{\Dl g(Q^2)}
\nc{\xq}{(x,Q^2)} \nc{\yq}{(y,Q^2)}
\nc{\Tt}{\tl{t}} \nc{\Ts}{\tl{s}} \nc{\Tu}{\tl{u}}
\nc{\Hs}{\ha{s}} \nc{\Ht}{\ha{t}} \nc{\Hu}{\ha{u}}
\nc{\Hsg}{\hat{\sg}}
\nc{\GeV}{\mbox{\rm GeV}}
\nc{\sS}{\!\not{\!s}}  
\nc{\pS}{\!\not{\!p}}  \nc{\kS}{\!\not{\!k}}
\nc{\poS}{\!\not{\!p}_1}  \nc{\pwS}{\!\not{\!p}_2}
\nc{\ptS}{\!\not{\!p}_3}  \nc{\pfS}{\!\not{\!p}_4}
\nc{\AS}{\!\not{\!\!A}}  \nc{\ASS}{\!\not{\!\!A}^*}
\nc{\BS}{\!\not{\!\!B}}  \nc{\BSS}{\!\not{\!\!B}^*}
\nc{\Tr}{\mbox{\rm Tr}}
\nc{\pT}{p_T}
\nc{\xT}{x_T}
\nc{\AoS}{\!\not{\!\!A}_1}  \nc{\AoSS}{\!\not{\!\!A}_1^*}
\nc{\AwS}{\!\not{\!\!A}_2}  \nc{\AwSS}{\!\not{\!\!A}_2^*}
\nc{\BoS}{\!\not{\!\!B}_1}  \nc{\BoSS}{\!\not{\!\!B}_1^*}
\nc{\BwS}{\!\not{\!\!B}_2}  \nc{\BwSS}{\!\not{\!\!B}_2^*}
\nc{\aS}{\!\not{\!a}}  \nc{\bS}{\!\not{\!b}}
\nc{\aoS}{\!\not{\!a}_1}  \nc{\boS}{\!\not{\!b}_1}
\nc{\awS}{\!\not{\!a}_2}  \nc{\bwS}{\!\not{\!b}_2}
\nc{\anS}{\!\not{\!a}_n}  \nc{\bnS}{\!\not{\!b}_n}
\nc{\anpoS}{\!\not{\!a}_{n+1}}  \nc{\bnpoS}{\!\not{\!b}_{n+1}}
\nc{\anmoS}{\!\not{\!a}_{n-1}}  \nc{\bnmoS}{\!\not{\!b}_{n-1}}
\nc{\refi}[1]{$^{\,\mbox{\scriptsize \ref{#1}}}$}
\nc{\refii}[2]{$^{\,\mbox{\scriptsize \ref{#1},\ref{#2}}}$}
\nc{\refiii}[3]{$^{\,\mbox{\scriptsize \ref{#1},\ref{#2},\ref{#3}}}$}
\nc{\refr}[2]{$^{\,\mbox{\scriptsize \ref{#1}--\ref{#2}}}$}
\nc{\sint}{\int \!\!}
\nc{\qB}{\stackrel{(-)}{q}}
\nc{\Lam}{\Lambda}
\nc{\noi}{\noindent}
\nc{\NL}{\newline}

\begin{document}
\title{Higgs Resonance Studies\\ At The First Muon Collider}

\author{Basim Kamal, William J.\ Marciano and Zohreh Parsa}
\address{Brookhaven National Laboratory, Upton, New York 11973}

\maketitle

\begin{abstract}
Higgs resonance signals and backgrounds at the First Muon Collider are
discussed. Effects due to beam polarization and background angular
distributions (forward-backward charge asymmetries) are examined.
The utility of those features for improving precision measurements
and narrow resonance ``discovery'' scans is described.
\end{abstract}

\section*{\mbox{}}

If the standard model Higgs boson has a mass $\lapp 160$ GeV
(i.e.\ below the $W^+W^-$ decay threshold), it will have a
very narrow width and can be resonantly
studied in the $s$-channel via 
$\mu^-\mu^+\RA H$ production at the First Muon Collider (FMC).
Within the framework of supersymmetry or  more general two Higgs
doublet scenarios, there can be several neutral spin zero bosons;
$h$, $H$, and $A$, all of which might be resonantly produced.
The lightest scalar, $h$, of supersymmetry is expected to be $\lapp 150$
GeV (and narrow), with the range 80--130 GeV favored. Precision 
electroweak measurements also tend to suggest,
via quantum loop sensitivity, a relatively light
Higgs. Hence, there are strong motivations to examine the capabilities
of the FMC for producing and studying relatively light scalar
resonances [1,2].

A strategy for ``light'' Higgs physics studies would be to first
discover the Higgs particle at LEPII, the Tevatron, or the LHC and then 
thoroughly scrutinize its properties on resonance at the FMC. There,
one would hope to precisely determine the Higgs mass, width, and
primary decay rates [3]. Besides those interesting physics 
studies, such an initiative would provide a nice
testing ground for muon collider technology and lay the foundation
for future much higher energy facilities.

The FMC Higgs resonance program would entail two stages:
1) ``Discovery'' via an energy scan which pinpoints the precise resonance
position and (perhaps) determines its width. Since pre-FMC efforts may only
determine the Higgs mass to ${\cal O}$(200 MeV) or worse and its width is 
expected to be narrow ${\cal O}$(1$\sim$ 30 MeV) for $m_H\lapp 160$ GeV,
the resonance scan may be very time consuming [3]. 2) Precision
measurements of the primary Higgs decay modes. Deviations from standard model 
expectations could point to additional Higgs structure or elucidate the
framework of supersymmetry [3]. (Expectations for $m_H=110$ GeV are 
illustrated in Table 1.)
\begin{table}
\caption{Expected signals and backgrounds (fully integrated) for a 
standard model Higgs with $m_H=110$ GeV, $\Gamma_H\simeq$ 3 MeV.
Muon collider resonance conditions with no polarization,
$\Dl E/E \simeq 3\times 10^{-5}$, and $L=0.05$ ${\rm fb}^{-1}$
are assumed. The total number of Higgs scalars produced
is $\sim 3000$. Realistic
efficiency and acceptance cuts are likely to dilute signal and
backgrounds for $b\B{b}$ and $c\B{c}$ by a $0.5$ factor.}
\label{table1}
\begin{tabular}{lddddd}
  $H\RA$ & \mbox{\hsp{25pt}} $b\B{b}$ \mbox{\hsp{25pt}} & 
\mbox{\hsp{25pt}} $c\B{c}$ \mbox{\hsp{25pt}} & 
\mbox{\hsp{25pt}} $\tau\B{\tau}$ \mbox{\hsp{25pt}} 
\\
\tableline
$N_S$ (events) &  $2400$  &  $210$ 
  &  $270$   \\
$N_B$ (events) &  $2520$  & 
  $2416$   &  $945$    \\
$\pm \sqrt{N_S+N_B}/N_S$  & \mbox{$\pm 0.03$}  & 
  \mbox{$\pm 0.24$}   &  \mbox{$\pm 0.13$}    \\
\end{tabular}
\end{table}

The Higgs resonance ``discovery'' capability and scan time
will depend on $N_S/\sqrt{N_B}$,
where $N_S$ is  the Higgs signal and $N_B$ is the expected background.
The precision measurement sensitivity will be determined by 
$N_S/\sqrt{N_B+N_S}$. For both, it will be extremely important to 
enhance the signal and suppress backgrounds as much as possible. To
that end, one should employ highly resolved $\mu^+\mu^-$ beams with
a very small energy spread. The proposed $\Dl E/E \simeq 3\times 10^{-5}$
is well matched to the narrow Higgs width. It allows $N_S/N_B \sim 
{\cal O}(1)$ for the primary $H\RA b\B{b}$ mode (see Table 1).
Unfortunately, high resolution is accompanied by luminosity loss. The
current goal of ${\cal L}_{\rm ave}\simeq 5\times 10^{30} {\rm cm}^{-2}
{\rm s}^{-1}$ on resonance is probably not ambitious enough.
One should strive for another order of magnitude in luminosity while
maintaining the outstanding beam energy resolution.

In this paper, we examine two additional ways of enhancing the Higgs
signal to background ratio: beam polarization and final state angular
distributions.  The Higgs signal $\mu^-\mu^+\RA H
\RA f\B{f}$ results from left-left (LL) or right-right (RR) beam 
polarizations and leads to an isotropic (i.e.\ constant) $f\B{f}$
signal in $\cos\theta$ (the angle between the $\mu^-$ and $f$).
Standard model backgrounds $\mu^-\mu^+\RA
\mbox{$\gm^*$ {\rm or} $Z^*$} \RA f\B{f}$ result from LR or RL
initial state polarizations and give rise to ($1+\cos^2\theta+\fr{8}{3}
A_{FB}\cos\theta$) angular distributions. Similar statements apply to
$WW^*$ and $ZZ^*$ final states, but those modes will not be discussed here.
%
%
\begin{figure}[t] 
\centerline{\epsfig{file=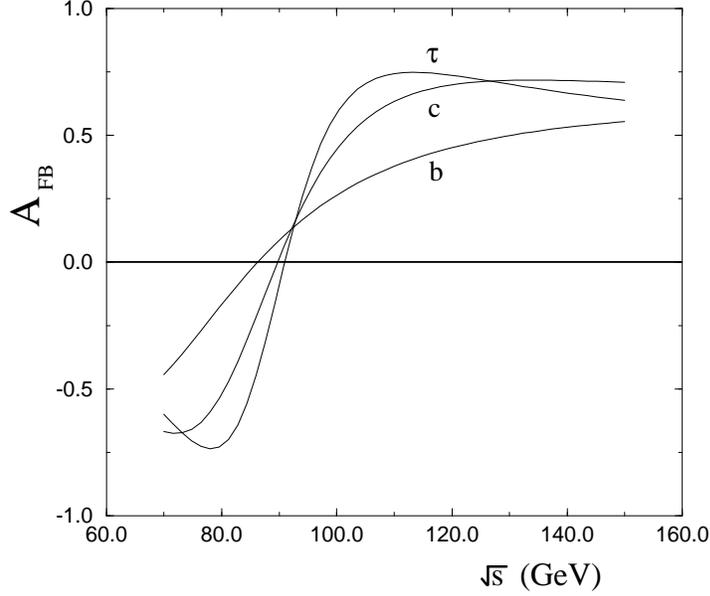,height=3.5in,width=4in}}
\vspace{10pt}
\caption{Forward-backward asymmetry for $\mu^-\mu^+\RA f\B{f}$.}
\label{fig1}
\end{figure}

To illustrate the difference between  signal, 
$\mu^-\mu^+\RA H \RA f\B{f}$, and background, $\mu^-\mu^+\RA
\mbox{$\gm^*$ {\rm or} $Z^*$} \RA f\B{f}$, we give the combined 
differential production rate with respect to $x\equiv\cos\theta=
4{\bf p_{\mu^-}}\C {\bf p_f}/s$ for polarized muon beams and fixed luminosity
\beqa
\label{e1}
\fr{dN(\mu^-\mu^+\RA f\B{f})}{dx} & = & \fr{1}{2} N_S (1 + P_+P_-)  \\
\nn & + &
\fr{3}{8} N_B [1-P_+P_- + (P_+-P_-)A_{LR}] 
  (1+x^2+\fr{8}{3} x A_{eff}).
\eeqa
$P_+(P_-)$ is the $\mu^+(\mu^-)$ polarization with $P=-1$ pure 
left-handed, $P=+1$ pure right handed, and $P=0$ unpolarized.
$N_S$ is the fully integrated ($-1<x\leq 1$) Higgs signal and $N_B$ the
integrated background for the case of unpolarized beams, $P_+=P_-=0$.
In that general expression,
\beq
A_{LR} \equiv \fr{\sg_{LR\RA LR}+\sg_{LR\RA RL} 
-\sg_{RL\RA RL}-\sg_{RL\RA LR}} 
{\sg_{LR\RA LR}+\sg_{LR\RA RL} +\sg_{RL\RA RL} + \sg_{RL\RA LR}},
\eeq
where, for example, $LR\RA LR$ stands for $\mu^-_L\mu^+_R\RA f_L\B{f}_R$.
The effective forward-backward asymmetry is given by
\beq
A_{eff} = \fr{A_{FB} + P_{eff}A_{LR}^{FB}}{1 + P_{eff}A_{LR}},
\eeq
with
\beqa
P_{eff} &=& \fr{P_+ - P_-}{1-P_+P_-}, \\
A_{FB} &=& \fr{3}{4}
\fr{\sg_{LR\RA LR}+\sg_{RL\RA RL} -\sg_{LR\RA RL}-\sg_{RL\RA LR}} 
{\sg_{LR\RA LR}+\sg_{RL\RA RL} +\sg_{LR\RA RL} + \sg_{RL\RA LR}}, \\
A_{LR}^{FB} &=& \fr{3}{4}
\fr{\sg_{LR\RA LR}+\sg_{RL\RA LR} -\sg_{LR\RA RL}-\sg_{RL\RA RL}} 
{\sg_{LR\RA LR}+\sg_{RL\RA LR} +\sg_{LR\RA RL} + \sg_{RL\RA RL}}.
\eeqa
and the $\mu^-_i\mu_j^+\RA f_{i'}\B{f}_{j'}$, cross sections ($i\neq j$)
are to lowest order
\beqa
\sg_{ij\RA i'j'} &=& (N_C) \sg_0 \lf[1-\fr{s}{m_W^2}\lf(1 +
T_{3\mu_i} - \fr{T_{3f_{i'}}}{Q_f}\lf(1+\fr{T_{3\mu_{i}}}{\sin^2\theta_W}
\rt)\rt)\rt]^2, \\
\nn T_{3\mu_L} &=& T_{3\tau_L} = T_{3b_L} = - T_{3c_L} =-1/2, \\
\nn T_{3f_R} &=& 0, \SSPP Q_\tau = 3Q_b = - \fr{3}{2} Q_c = -1 
\SSP (N_C = 3 \text{ for } f=b,c).
\eeqa
Realistic cuts, efficiencies, systematic errors etc, will not
be considered. They are likely to dilute the $b\B{b}$ and $c\B{c}$
event rates by a factor of 0.5.
In addition, we ignore the radiative $Z$ production
tail under the assumption such events are vetoed.

\begin{figure}[t] 
\centerline{\epsfig{file=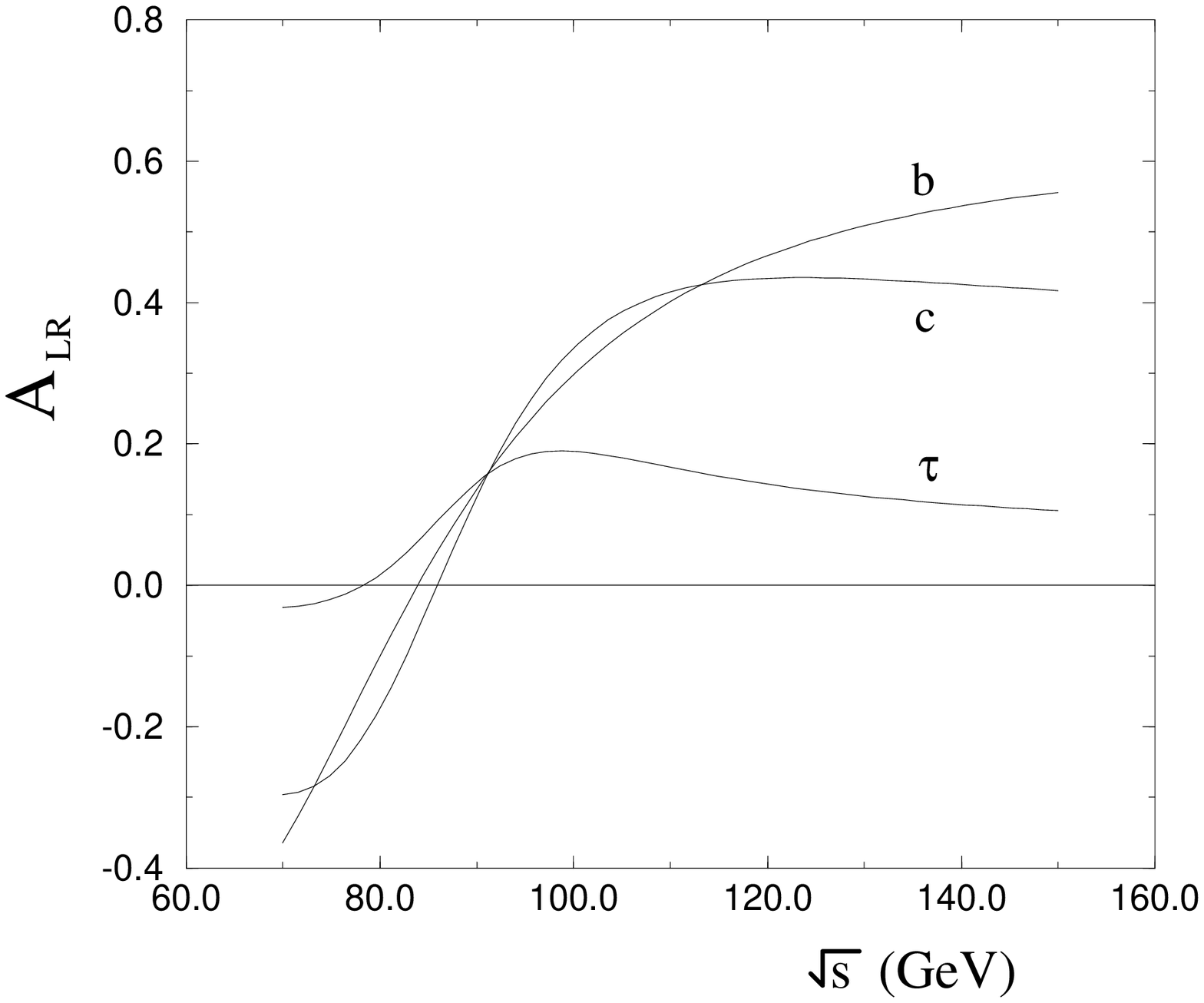,height=3.5in,width=4in}}
\vspace{10pt}
\caption{Left-right asymmetry for $\mu^-\mu^+\RA f\B{f}$.}
\label{fig2}
\end{figure}
The (unpolarized) forward-backward asymmetries are illustrated in 
Fig.\ 1. Note that $A_{FB}$ is large (near maximal) for $\tau\B{\tau}$
and $c\B{c}$ in the region of interest. As we shall see, that feature can 
help in 
discriminating signal from background.

In principle, large polarization can be important for enhancing
``discovery'' and precision measurement sensitivity for the Higgs.
From Eq.\ (\ref{e1}), we find that
$N_S/\sqrt{N_B}$ is  enhanced (
for integrated signal and background) by the factor
\beq
\kappa_{\rm pol} = \fr{1+P_+P_-}{\sqrt{1-P_+P_-+(P_+-P_-)A_{LR}}}\,\, ,
\eeq
where the $A_{LR}$ are shown in Fig.\ 2.
That result generalizes the $P_+=P_-$ case [4]. For natural
beam polarization [1], $P_+=P_-=0.2$ (assuming spin rotation of
one beam), the enhancement factor is only 1.06. For larger polarization,
$P_+=P_-=0.5$, one obtains a 1.44 enhancement factor (statistically
equivalent to about a factor of 2 luminosity increase). Unfortunately,
obtaining 0.5 polarization simply by muon energy cuts reduces each
beam intensity [1] by a factor of $1/4$, resulting in a luminosity
reduction by $1/16$. Such a tradeoff is clearly unacceptable. Polarization
will be a useful tool in Higgs resonance studies only if high polarization
is achievable with little luminosity loss.
Tau final state polarizations can also be used to help improve the
$H\RA \tau\B{\tau}$ measurement, but will not be discussed here.

\begin{figure}[t] 
\centerline{\epsfig{file=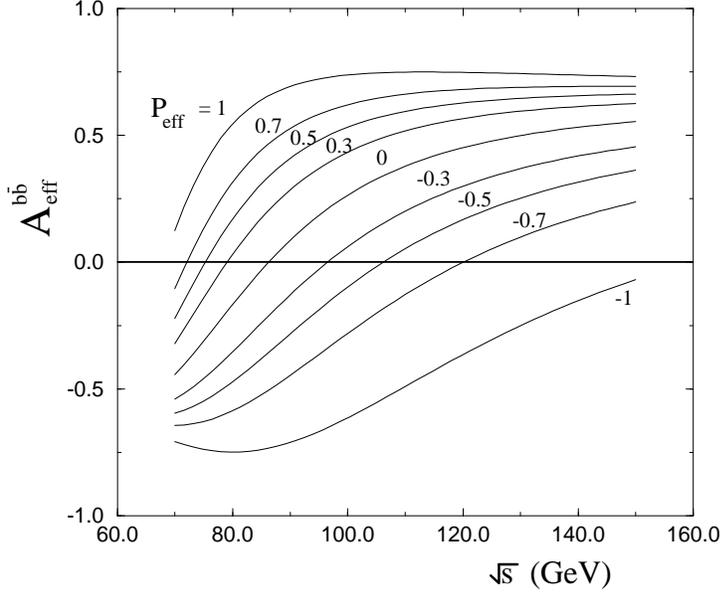,height=3.5in,width=4in}}
\vspace{10pt}
\caption{Effective forward-backward asymmetry for $\mu^-\mu^+\RA b\B{b}$.}
\label{fig3}
\end{figure}
Some ``discovery'' or sensitivity enhancement can also be obtained from
angular discrimination. A proper study would include detector acceptance
cuts and maximum likelihood fits. Here, we wish to only crudely 
approximate the gain. For that purpose, we assume perfect (infinitesimal)
binning and obtain the measurement sensitivity enhancement factor
\beq
\label{e8}
\fr{1}{2} (1+P_+P_-) \sqrt{N_S+N_B} \lf[ \int \fr{dx}{dN/dx}\rt]^{1/2},
\eeq
which becomes, from Eq.\ (\ref{e1}),
\beq
\kappa_{\rm pol} 
\sqrt{\fr{2}{3}} \sqrt{\fr{N_S+N_B}{N_B}} \lf( \fr{\tan^{-1}\lf(
\fr{2}{\zeta}
\sqrt{1-\fr{16}{9}A_{eff}^2+\zeta}\rt)}{\sqrt{1-\fr{16}{9}A_{eff}^2
+\zeta}}
\rt)^{1/2} \!\!\!, \SSPP \zeta\equiv \fr{4}{3} \fr{N_S}{N_B} 
\fr{\kappa_{\rm pol}^2}{1+P_+P_-}\,\,.
\eeq
For $A_{eff}
\simeq 3/4$, $\zeta\simeq 0.38$ (which roughly applies to 
$\tau\B{\tau}$) and $P_+=P_-=0$, one finds a sensitivity enhancement of 1.33. 
That means the $\pm 13$\% statistical error in Table 1 would be
reduced to $\pm 10$\%.
Similar 
sensitivity enhancements apply to $c\B{c}$. In the case of $H\RA b\B{b}$, the 
primary discovery mode, $A_{eff}\simeq 0.4$ and one finds only a
3\% enhancement. One can increase the effective $b\B{b}$ 
forward-backward asymmetry via polarization (see Fig.\ 3). However,
one must again confront the issue of luminosity loss. 

In the case of ``discovery'', a very large forward-backward asymmetry
(near maximal) can, in principle, significantly reduce the scan time.
For the highly idealized  coverage and binning assumed above, the time 
is reduced by the factor
\beq
\fr{1}{\kappa_{\rm pol}^2} \fr{3}{\pi}
\sqrt{1-\fr{16}{9} A_{eff}^2}\,\,.
\eeq
Of course, that naive formula must be corrected for realistic
acceptances, efficiencies, etc.; so, it should not be taken too
literally (particularly for $A_{eff}\simeq 3/4$). Nevertheless, applying
it to the $b\B{b}$ discovery mode with ``natural'' $P_+=P_-=0.2$
and $A_{eff}\simeq 0.37$ gives a scan reduction time factor of $0.74$.

The $H\RA \tau\B{\tau}$ ``discovery'' time is about 15 times longer than
that of
the $b\B{b}$ (with efficiencies) for fully integrated signals. Employing
$A_{FB}\simeq 0.743$ and assuming tau detection down to about $15^o$ from
the beams, reduces that time by about a factor of $6\sim 7$, making it somewhat
less than 1/2 as effective as $b\B{b}$. Using both together along with
all background angular information should, therefore, reduce the scan
time by almost a factor of 2 compared to using the integrated $b\B{b}$
signal alone. Such a reduction would be extremely welcome, particularly
if the luminosity is less than expected.

In conclusion, we have shown that polarization is 
potentially useful for Higgs
resonance studies, but only if the 
accompanying luminosity reduction is not significant. Large 
forward-backward asymmetries can  also be used to enhance the Higgs
``discovery'' signal or improve precision measurements, particularly
for $\tau\B{\tau}$. However, to
make the $s$-channel Higgs ``factory'' a compelling facility, one
must focus on attaining the 
outstanding beam resolution 
assumed here and maintaining the
highest luminosity possible.

\end{document}